\documentclass[10pt]{article}
\usepackage{graphicx}
\usepackage{amsmath}
\usepackage{amssymb}
\usepackage{caption2}
\begin{document}
\begin{center}
{\large {\bf \sc{  Analysis of the mass and width of the  $X^*(3860)$  with QCD sum rules
  }}} \\[2mm]
Zhi-Gang  Wang \footnote{E-mail: zgwang@aliyun.com.  }   \\
 Department of Physics, North China Electric Power University, Baoding 071003, P. R. China
\end{center}

\begin{abstract}
In this article, we tentatively assign the $X^*(3860)$ to be  the $C\gamma_5\otimes \gamma_5C$ type scalar  tetraquark  state,   study its mass and width with the QCD sum rules, special attention is paid to calculating the hadronic coupling constants $G_{X\eta_c\pi}$ and $G_{XDD}$ concerning the tetraquark state. We obtain the values $M_{X}=3.86 \pm 0.09\,\rm{GeV}$ and  $\Gamma_{X}= 202\pm 146 \,\rm{MeV}$, which are consistent with the experimental data.   The numerical result  supports assigning  the   $X^*(3860)$ to be  the $C\gamma_5\otimes \gamma_5C$ type scalar  tetraquark  state.
\end{abstract}

PACS number: 12.39.Mk, 12.38.Lg

Key words: Tetraquark  state, QCD sum rules

\section{Introduction}

Recently, the Belle collaboration  performed a full amplitude analysis of the process $e^+ e^- \rightarrow
J/\psi D \bar{D}$ based on the $980 \rm{fb}^{-1}$   data sample collected by the Belle detector at the asymmetric-energy $e^+ e^-$   collider KEKB, and observed
   a new charmoniumlike state $X^*(3860)$  that decays to $D \bar{D}$ with  a significance of $6.5\sigma$, the measured mass  is
   $3862^{+26}_{-32}{}^{+40}_{-13}\,\rm{MeV}$  and width is $201^{+154}_{-67}{}^{+88}_{-82}\,\rm{MeV}$ \cite{Belle-3860}. The $J^{PC} =0^{++}$   hypothesis is favored over the $2^{++}$   hypothesis at the level of $2.5\sigma$. The Belle collaboration assigned the $X^*(3860)$ in stead of the $X(3915)$ to be the $\chi_{c0}(\rm 2P)$ state   \cite{Belle-3860}. The mass of the state  $\chi_{c0}(\rm 2P)$  from the non-relativistic potential model,  the Godfrey-Isgur relativized potential model and
   the screened potential model is $3852\,\rm{MeV}$, $3916\,\rm{MeV}$ and $3842\,\rm{MeV}$, respectively \cite{GI-model,ChaoKT}.

In 2004, the  Belle collaboration  observed  the $X(3915)$  in the $\omega J/\psi$  mass spectrum in the
 exclusive $B \to K \omega J/\psi$ decays \cite{Belle2004}. In 2007, the BaBar collaboration confirmed the $X(3915)$ in the $\omega J/\psi$  mass spectrum in the  exclusive $B \to K \omega J/\psi$ decays \cite{BaBar2007}. In 2010, the   Belle collaboration confirmed the $X(3915)$    in the two-photon process $\gamma\gamma\to \omega J/\psi$ \cite{Belle2010}.

 In Ref.\cite{Lebed-3915},  Lebed and Polosa  propose that the  $X(3915)$ is the lightest
 $cs\bar{c} \bar{s}$ scalar tetraquark state based on  lacking of the observed $D\bar D$ and $D^*\bar{D}^*$
decay modes, and attribute  the single known decay mode $J/\psi \omega$ to the $\omega-\phi$ mixing effect.
In Refs.\cite{Wang-3915-CgmCgm,Wang-3915-C5C5}, we study the $C\gamma_\mu\otimes \gamma^\mu C$-type, $C\gamma_\mu\gamma_5\otimes \gamma_5\gamma^\mu C$-type,
$C\gamma_5\otimes \gamma_5 C$-type, $C\otimes  C$-type $cs\bar{c}\bar{s}$ scalar tetraquark states with the QCD sum rules in a systematic way, and obtain the predictions $M_{C\gamma_\mu\otimes \gamma^\mu C}=3.92^{+0.19}_{-0.18}\,\rm{GeV}$ and $M_{C\gamma_5\otimes \gamma_5C}=3.89\pm0.05\,\rm{GeV}$, which support assigning the $X(3915)$ to be the $C\gamma_\mu\otimes \gamma^\mu C$-type or $C\gamma_5\otimes \gamma_5 C$-type $cs\bar{c}\bar{s}$ scalar tetraquark state.

Naively, we expect the  $SU(3)$ breaking effect is about $m_s-m_q=135\,\rm{MeV}$, while  the QCD sum rules indicate that the mass gaps  $M_{cs\bar{c}\bar{s}}-M_{cq\bar{c}\bar{q}}$ are less than or much less than $90\,\rm{MeV}$ for the scalar, vector, axialvector diquark-antidiquark type hidden-charm  tetraquark states \cite{WangScalar-2009,Wang-Axial-V-tetraquark,Wang-4660-2014}. If the $SU(3)$ breaking effects are small indeed for the diquark-antidiquark type hidden-charm  tetraquark states, the $X^*(3860)$ and $X(3915)$ can be assigned to be the scalar tetraquark states with the symbolic quark structures $\bar{c}c \frac{\bar{u}u+\bar{d}d}{\sqrt{2}}$ and $\bar{c}c\bar{s}s$, respectively.   In Ref.\cite{Wang-Scalar-MPLA}, we study the lowest $C\gamma_5\otimes \gamma_5C$ type  scalar  hidden-charm tetraquark state with  the QCD sum rules and obtain the mass $M=\left(3.82^{+0.08}_{-0.08}\right)\,\rm{GeV}$, which is consistent with the value from the Belle collaboration  \cite{Belle-3860}.

In Ref.\cite{Wang-1601}, we update the value of the effective $c$-quark mass ${\mathbb{M}}_c$ in determining the optimal energy scales of the QCD spectral densities in the QCD sum rules  for the hidden-charm tetraquark states by the  empirical formula  $\mu=\sqrt{M^2_{X/Y/Z}-(2{\mathbb{M}}_c)^2}$, where the $X$, $Y$, $Z$ denote the tetraquark states.
So the predicted mass  of the $C\gamma_5\otimes \gamma_5C$ type     hidden-charm tetraquark state in Ref.\cite{Wang-Scalar-MPLA} should be updated.
In Ref.\cite{Wang-Scalar-MPLA}, we take the old value ${\mathbb{M}}_c=1.80\,\rm{GeV}$, now we take the updated value ${\mathbb{M}}_c=1.82\,\rm{GeV}$ \cite{Wang-1601}, and expect to extract a slightly different  mass $M_X$  at a slightly different energy scale $\mu$ in a consistent way according to the energy scale formula $\mu=\sqrt{M^2_{Z}-(2{\mathbb{M}}_c)^2}$. Variations of the energy scales  $\mu$ lead to changes  of integral range $4m_c^2(\mu)-s_0$ of the variable  $ds$ besides the QCD spectral density  $\rho(s)$ (See Eq.(4) in Sec.2),
therefore change of the Borel window and predicted mass and pole residue.
Moreover, it is interesting to study the decay widths of the tetraquark states with the QCD sum rules by taking into account all the Feynman diagrams \cite{Wang-4200,Zhu-4200-Wang-5568} instead of  only the connected Feynman diagrams \cite{Wang-Scalar-MPLA,Nielsen3900}. Furthermore,  the  over simplified hadron representation  chosen in Ref.\cite{Wang-Scalar-MPLA} should be modified. In this article, we assign the $X^*(3860)$ to be the $C\gamma_5\otimes \gamma_5C$ type  scalar  hidden-charm tetraquark state, and restudy its mass and width with the QCD sum rules in details.

The article is arranged as follows:  we derive the QCD sum rules for
the mass and width of the   $X^*(3860)$  in section 2 and section 3 respectively; section 4 is reserved for our conclusion.

\section{The mass of the $C\gamma_5\otimes \gamma_5C$ type  scalar  hidden-charm tetraquark state }
In the following, we write down  the two-point correlation function $\Pi(p)$  in the QCD sum rules,
\begin{eqnarray}
\Pi(p)&=&i\int d^4x e^{ip \cdot x} \langle0|T\left\{J(x) J^{\dagger}(0)\right\}|0\rangle \, ,
\end{eqnarray}
where
\begin{eqnarray}
J(x)&=&\varepsilon^{ijk}\varepsilon^{imn} u^j(x)C\gamma_5c^k(x) \bar{d}^m(x)\gamma_5 C \bar{c}^n(x)  \, ,
\end{eqnarray}
 the $i$, $j$, $k$, $m$, $n$ are color indexes, the $C$ is the charge conjunction matrix. We choose  the   current $J(x)$ to interpolate the
   tetraquark state $X^*(3860)$ (to be more precise, the charged partner of the $X^*(3860)$, they have degenerate masses in the isospin limit).

At the phenomenological side,  we insert  a complete set of intermediate hadronic states with
the same quantum numbers as the current operator $J(x)$ into the
correlation function $\Pi(p)$  to obtain the hadronic representation
\cite{SVZ79,Reinders85}, and isolate the ground state
contribution,
\begin{eqnarray}
\Pi(p)&=&\frac{\lambda_{X}^2}{M^2_{X}-p^2}  +\cdots  \, ,
\end{eqnarray}
where the pole residue  $\lambda_{X}$ is defined by $ \langle 0|J (0)|X^*(3860)\rangle=\lambda_{X}$.

We carry out the
operator product expansion to the vacuum condensates  up to dimension-10, and obtain the QCD spectral density through dispersion relation, then
 we take the
quark-hadron duality and perform Borel transform  with respect to
the variable $P^2=-p^2$ to obtain  the following QCD sum rule,
\begin{eqnarray}
\lambda^2_{X}\, \exp\left(-\frac{M^2_{X}}{T^2}\right)= \int_{4m_c^2}^{s_0} ds\, \rho(s) \, \exp\left(-\frac{s}{T^2}\right) \, ,
\end{eqnarray}
where the $T^2$ is the Borel parameter and the $s_0$ is the continuum threshold parameter.
The explicit expression of the QCD spectral density $\rho(s)$ is presented in Refs.\cite{Wang-3915-C5C5,Wang-Scalar-MPLA}.

 We derive  Eq.(4) with respect to  $\tau=\frac{1}{T^2}$, then eliminate the
 pole residue  $\lambda_{X}$ to obtain the QCD sum rule for the mass,
 \begin{eqnarray}
 M^2_{X}= \frac{-\frac{d}{d \tau } \int_{4m_c^2}^{s_0} ds\,\rho(s)\,e^{-\tau s}}{\int_{4m_c^2}^{s_0} ds \,\rho(s)\,e^{-\tau s}}\, .
\end{eqnarray}

We take  the standard values of the vacuum condensates $\langle
\bar{q}q \rangle=-(0.24\pm 0.01\, \rm{GeV})^3$,   $\langle
\bar{q}g_s\sigma G q \rangle=m_0^2\langle \bar{q}q \rangle$,
$m_0^2=(0.8 \pm 0.1)\,\rm{GeV}^2$, $\langle \frac{\alpha_s
GG}{\pi}\rangle=0.012\pm0.003\,\rm{GeV}^4 $    at the energy scale  $\mu=1\, \rm{GeV}$
\cite{SVZ79,Reinders85,Colangelo-Review}, and choose the $\overline{MS}$ mass $m_{c}(m_c)=(1.275\pm0.025)\,\rm{GeV}$  from the Particle Data Group \cite{PDG}.
Moreover, we take into account the energy-scale dependence of  the input parameters,
\begin{eqnarray}
\langle\bar{q}q \rangle(\mu)&=&\langle\bar{q}q \rangle(Q)\left[\frac{\alpha_{s}(Q)}{\alpha_{s}(\mu)}\right]^{\frac{4}{9}}\, ,\nonumber\\
 \langle\bar{q}g_s \sigma Gq \rangle(\mu)&=&\langle\bar{q}g_s \sigma Gq \rangle(Q)\left[\frac{\alpha_{s}(Q)}{\alpha_{s}(\mu)}\right]^{\frac{2}{27}}\, ,\nonumber\\
m_c(\mu)&=&m_c(m_c)\left[\frac{\alpha_{s}(\mu)}{\alpha_{s}(m_c)}\right]^{\frac{12}{25}} \, ,\nonumber\\
\alpha_s(\mu)&=&\frac{1}{b_0t}\left[1-\frac{b_1}{b_0^2}\frac{\log t}{t} +\frac{b_1^2(\log^2{t}-\log{t}-1)+b_0b_2}{b_0^4t^2}\right]\, ,
\end{eqnarray}
  where $t=\log \frac{\mu^2}{\Lambda^2}$, $b_0=\frac{33-2n_f}{12\pi}$, $b_1=\frac{153-19n_f}{24\pi^2}$, $b_2=\frac{2857-\frac{5033}{9}n_f+\frac{325}{27}n_f^2}{128\pi^3}$,  $\Lambda=213\,\rm{MeV}$, $296\,\rm{MeV}$  and  $339\,\rm{MeV}$ for the flavors  $n_f=5$, $4$ and $3$, respectively  \cite{PDG}.
We  tentatively take the continuum threshold parameter  to be  $\sqrt{s_0}=(4.4\pm 0.1)\,\rm{GeV}$, i.e. $\sqrt{s_0}=M_{X}+(0.4-0.6)\,\rm{GeV}$. In the  scenario of tetraquark  states, the QCD sum rules indicate that the $Z_c(3900)$ and $Z(4430)$ can be tentatively assigned to be the ground state and the first radial excited state of the axialvector tetraquark states, respectively \cite{Wang4430}, the $X(3915)$ and $X(4500)$ can be tentatively assigned to be the ground state and the first radial excited state of the scalar tetraquark states, respectively \cite{Wang-3915-CgmCgm,Wang-3915-C5C5}. The energy gap between the ground state and the first radial excited state of the hidden-charm tetraquark states is about $0.6\,\rm{GeV}$.

In Refs.\cite{Wang-4660-2014,WangHuangTao-3900,Wang-Huang-NPA-2014}, we study the acceptable energy scales of the QCD spectral densities  for the hidden-charm (hidden-bottom) tetraquark states  in the QCD sum rules in details for the first time,  and suggest an  empirical formula  $\mu=\sqrt{M^2_{X/Y/Z}-(2{\mathbb{M}}_Q)^2}$ to determine  the optimal  energy scales, where the $X$, $Y$, $Z$ denote the tetraquark states, and the ${\mathbb{M}}_Q$ denotes the effective heavy quark masses. The energy scale formula works well for the  $X(3872)$, $Z_c(3900)$,  $X(3915)$, $Z_c(4020/4025)$, $Y(4140)$, $Z(4430)$, $X(4500)$, $Y(4660)$, $X(4700)$, $Z_b(10610)$  and $Z_b(10650)$.  In Ref.\cite{Wang-Scalar-MPLA}, we choose the old value ${\mathbb{M}}_c=1.80\,\rm{GeV}$ to study the mass of the lowest scalar hidden-charm tetraquark state. In this article, we choose the updated value ${\mathbb{M}}_c=1.82\,\rm{GeV}$ \cite{Wang-1601}, and obtain the optimal energy scale $\mu=1.3\,\rm{GeV}$  for the QCD spectral density, the prediction is changed slightly. In fact, the empirical energy scale formula  $\mu=\sqrt{M^2_{X/Y/Z}-(2{\mathbb{M}}_c)^2}$ serves as a constraint to obey.

We search for the optimal Borel parameter to satisfy the two criteria (pole dominance and convergence of the operator product
expansion) of the QCD sum rules, and obtain the value $T^2=(2.5-2.9)\,\rm{GeV}^2$. In Fig.1, we plot the pole contribution with variations of the Borel parameter $T^2$,
 the pole contribution is about $(46-70)\%$ in the Borel window between the two vertical lines. In Fig.2, we plot the contributions of  different terms in the operator product expansion    with variations  of the Borel parameter $T^2$ for the central value of the continuum threshold parameter $s_0$. In the Borel window, the main contributions come from the vacuum condensates of dimensions $0$, $3$, $5$ and $6$,
the contributions of the vacuum condensates of dimensions 8 and 10 are about $-(3-6)\%$ and $<1\%$, respectively. The two criteria of the QCD sum rules are fully satisfied, we expect to make reliable prediction.

\begin{figure}
 \centering
 \includegraphics[totalheight=7cm,width=10cm]{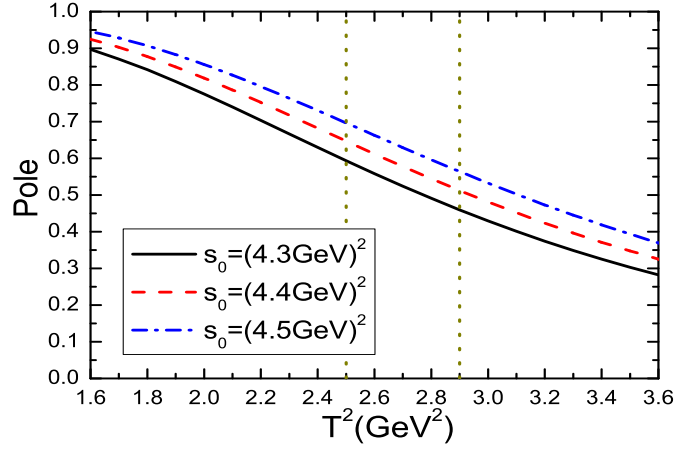}
          \caption{ The pole contribution with variations  of the Borel parameter $T^2$.  }
\end{figure}

\begin{figure}
 \centering
 \includegraphics[totalheight=7cm,width=10cm]{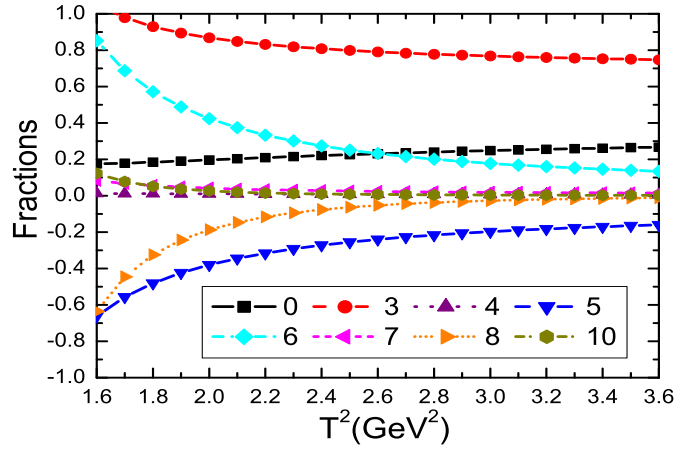}
          \caption{ The contributions of  different terms in the operator product expansion    with variations  of the Borel parameter $T^2$, where the $0$, $3$, $4$, $5$, $6$, $7$, $8$ and $10$ denote the dimensions of the vacuum condensates.  }
\end{figure}

We take  into account all uncertainties of the input parameters,
and obtain the values of the mass and pole residue of
 the    $X^*(3860)$, which are  shown explicitly in Fig.3,
\begin{eqnarray}
M_{X}&=&3.86 \pm 0.09\,\rm{GeV} \, ,  \nonumber\\
\lambda_{X}&=&(2.02\pm0.34)\times 10^{-2}\,\rm{GeV}^5 \,   .
\end{eqnarray}
The predicted mass $M_{X}=3.86 \pm 0.09\,\rm{GeV}$  is in excellent agreement  with the experimental value $3862^{+26}_{-32}{}^{+40}_{-13}\,\rm{MeV}$ within uncertainties \cite{Belle-3860}.  The QCD sum rules favors  assigning the $X^*(3860)$ to be  the $C\gamma_5\otimes\gamma_5C$ type hidden-charm tetraquark state. However, the assignment of the $X(3915)$ as the $C\gamma_5\otimes\gamma_5C$ type hidden-charm tetraquark state with the symbolic structure $\bar{c}c \frac{\bar{u}u+\bar{d}d}{\sqrt{2}}$  is not excluded, as the predicted mass $M_{X}=3.86 \pm 0.09\,\rm{GeV}$ is also compatible with the experimental value $3918.4\pm 1.9\,\rm{MeV}$ of the mass of the $X(3915)$ within uncertainty \cite{PDG}. We can study the width to obtain more reliable assignment.  On the other hand, the Belle  collaboration observed the $X(3940)$ in the decays to the meson pair $D^*\bar{D}$ \cite{Belle-X3940}, absence of the decays $X(3940) \to D\bar{D}$ indicates the favored  quantum numbers of the $X(3940)$ are $J^{PC}=0^{-+}$, which differ from the quantum numbers $J^{PC}=0^{++}$ of the interpolating current $J(x)$.

 \begin{figure}
 \centering
 \includegraphics[totalheight=5cm,width=7cm]{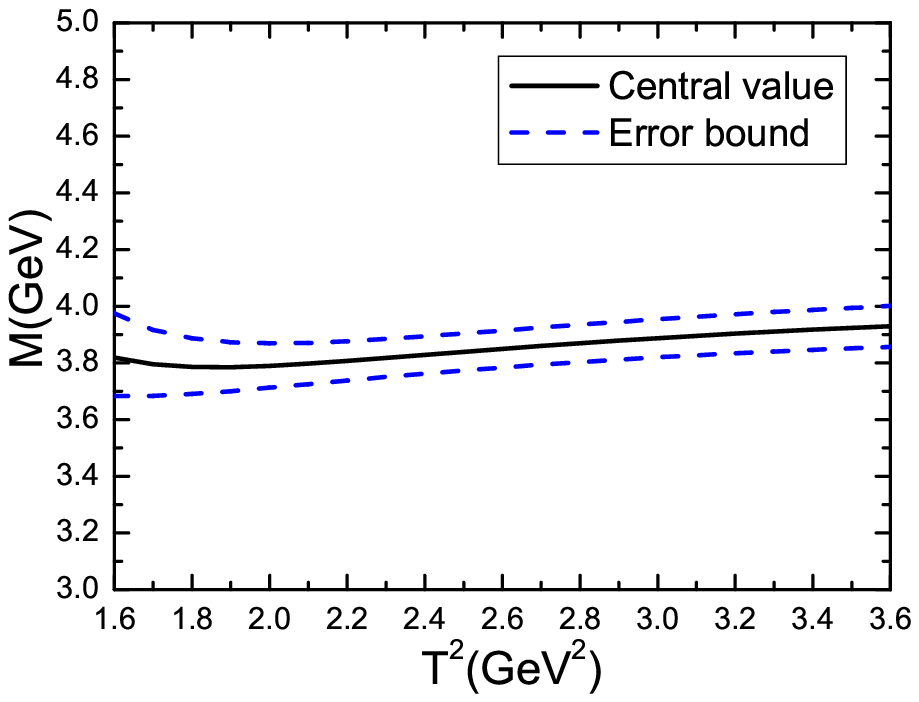}
 \includegraphics[totalheight=5cm,width=7cm]{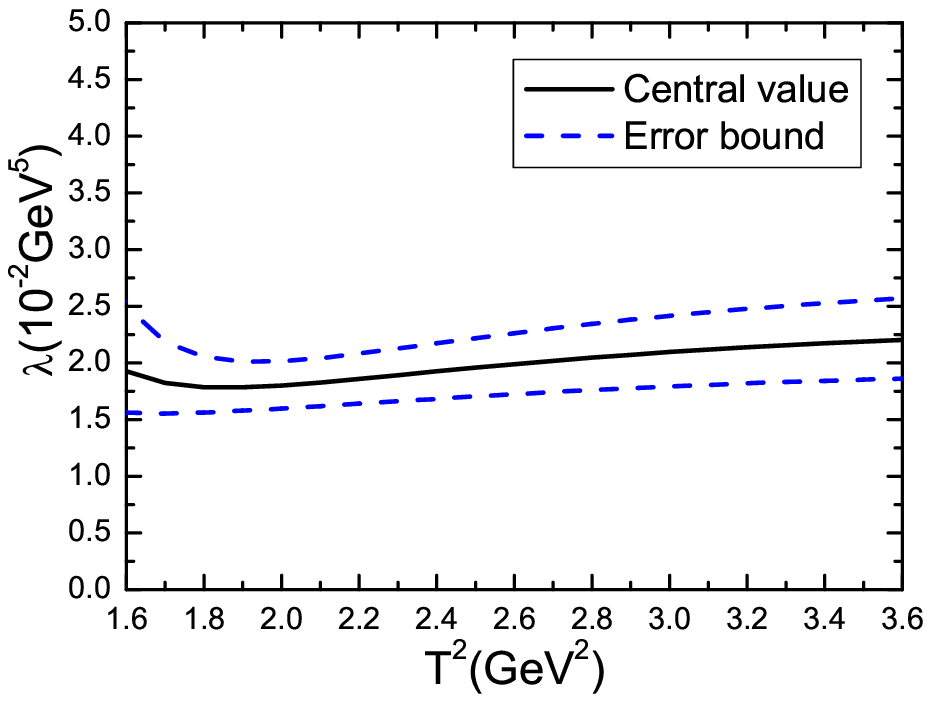}
        \caption{ The mass and pole residue  of the $X^*(3860)$  with variations  of the Borel parameter $T^2$.  }
\end{figure}

\section{The width of the $X^*(3860)$ as scalar tetraquark state }

We study the two-body strong decays $X^*(3860)\to \eta_c\pi^{-}$ and $D^-D^0$  with the following three-point correlation functions
$\Pi_{1}(p,q)$ and $\Pi_{2}(p,q)$, respectively,
\begin{eqnarray}
\Pi_{1}(p,q)&=&i^2\int d^4xd^4y e^{ip \cdot x}e^{iq \cdot y}\langle 0|T\left\{J_{\eta_c}(x)J_{\pi}(y)J(0)\right\}|0\rangle\, ,   \\
\Pi_{2}(p,q)&=&i^2\int d^4xd^4y e^{ip\cdot x}e^{iq\cdot y}\langle 0|T\left\{J_{D}(x)J_{D}(y)J(0)\right\}|0\rangle \, ,
\end{eqnarray}
where the currents
\begin{eqnarray}
J_{\eta_c}(x)&=&\bar{c}(x)i\gamma_5 c(x) \, ,\nonumber \\
J_5^{\pi}(y)&=&\bar{u}(y)i\gamma_5 d(y) \, ,  \\
J_{D}(x)&=&\bar{c}(x)i\gamma_5 d(x) \, ,\nonumber \\
J_{D}(y)&=&\bar{u}(y)i\gamma_5 c(y) \, ,
\end{eqnarray}
interpolate the mesons $\eta_c$, $\pi^-$,  $D^-$ and $D^0$, respectively.

At the QCD side, the correlation functions $\Pi_{1}(p,q)$ and $\Pi_{2}(p,q)$ can be written as
\begin{eqnarray}
\Pi_{1}(p,q)&=& \int_{4m_c^2}^{s_0} ds \int_{0}^{u_0} du  \frac{\rho_1(s,u)}{(s-p^2)(u-q^2)}+\int_{4m_c^2}^{s_0} ds \int_{u_0}^\infty du  \frac{\rho_1(s,u)}{(s-p^2)(u-q^2)} \nonumber\\
 &&+\int_{s_0}^{\infty} ds \int_{0}^{u_0} du  \frac{\rho_1(s,u)}{(s-p^2)(u-q^2)}+\int_{s_0}^\infty ds \int_{u_0}^\infty du  \frac{\rho_1(s,u)}{(s-p^2)(u-q^2)} \, ,\\
\Pi_{2}(p,q)&=& \int_{m_c^2}^{s_0} ds \int_{m_c^2}^{u_0} du  \frac{\rho_2(s,u)}{(s-p^2)(u-q^2)}+ \int_{m_c^2}^{s_0} ds\int_{u_0}^\infty du  \frac{\rho_2(s,u)}{(s-p^2)(u-q^2)} \nonumber\\
&&+ \int_{s_0}^\infty ds\int_{m_c^2}^{u_0} du  \frac{\rho_2(s,u)}{(s-p^2)(u-q^2)}+\int_{s_0}^\infty ds \int_{u_0}^\infty du  \frac{\rho_2(s,u)}{(s-p^2)(u-q^2)} \, ,
\end{eqnarray}
where the $\rho_{1/2}(s,u)$ are the QCD spectral densities, the $s_0$ and $u_0$ are the continuum threshold parameters. The QCD spectral densities $\rho_{1}(s,u)$ and $\rho_{2}(s,u)$ are independent on the  $(p\pm q)^2$ except for  some non-singular terms $p\cdot q$, $(p\cdot q)^2$, etc, the variables $ds$ and $du$ are independent,
which differ from the  QCD spectral densities in the QCD sum rules for the  hadronic coupling constants $G_{\Lambda_cND}$, $G_{\Lambda_bNB}$, $G_{\Sigma_cND}$, $G_{\Sigma_bNB}$, $G_{\Lambda_cND^*}$, $G_{\Lambda_bNB^*}$, $G_{\Sigma_cND^*}$, $G_{\Sigma_bNB^*}$ \cite{Azizi-2015,Azizi-2015-YGL}, $ G_{B_c^*B_c\Upsilon}$, $ G_{B_c^*B_c J/\psi}$, $ G_{B_cB_c\Upsilon}$, $ G_{B_cB_c J/\psi}$, $G_{D_2^*D\pi}$, $G_{D_{s2}^*DK}$, $G_{B_2^*B\pi}$, $G_{B_{s2}^*BK}$ \cite{WangPRD-2014}, in those  case the QCD spectral densities depend on the $(p\pm q)^2$ explicitly, the variables $ds$ and $du$ should obey  special constraints among  the $s$, $u$ and $(p\pm q)^2$ according to dispersion relations or Cutkosky's rules \cite{WangPRD-2014}.  The strong decays $X^*(3860)\to \eta_c\pi^{-}$ and $D^-D^0$ take place through fall-apart mechanism, no quark-antiquark pair is created from the vacuum, which differs  from the two-body strong decays of the conventional mesons and baryons significantly.

At the hadronic side, we insert  a complete set of intermediate hadronic states with
the same quantum numbers as the current operators into the three-point
correlation functions $\Pi_{1}(p,q)$, $\Pi_{2}(p,q)$ and  isolate the ground state
contributions to obtain the following results,
\begin{eqnarray}
\Pi_{1}(p,q)&=& \frac{f_{\eta_c}M_{\eta_c}^2f_{\pi}M_{\pi}^2\lambda_{X}G_{X\eta_c \pi}}{2m_c(m_u+m_d)} \frac{1}{(M_{X}^2-p^{\prime2})(M_{\eta_c}^2-p^2)(M_{\pi}^2-q^2)}\nonumber\\
&&+ \frac{1}{(M_{X}^2-p^{\prime2})(M_{\eta_c}^2-p^2)} \int_{s^0_\pi}^\infty dt\frac{\rho_{X\pi}(p^2,t,p^{\prime 2})}{t-q^2}\nonumber\\
&&+ \frac{1}{(M_{X}^2-p^{\prime2})(M_{\pi}^2-q^2)} \int_{s^0_{\eta_c}}^\infty dt\frac{\rho_{X\eta_c}(t,q^2,p^{\prime 2})}{t-p^2}  \nonumber\\
&&+ \frac{1}{(M_{\eta_c}^2-p^2)(M_{\pi}^2-q^2)} \int_{s^0_X}^\infty dt\frac{\rho_{X\eta_c}(p^2,q^2,t)+\rho_{X\pi}(p^2,q^2,t)}{t-p^{\prime2}}+\cdots \, ,
\end{eqnarray}
\begin{eqnarray}
\Pi_{2}(p,q)&=& \frac{f_{D}^2M_{D}^4 \lambda_{X}G_{XD D}}{4m_c^2} \frac{1}{(M_{X}^2-p^{\prime2})(M_{D^-}^2-p^2)(M_{D^0}^2-q^2)}\nonumber\\
&&+ \frac{1}{(M_{X}^2-p^{\prime2})(M_{D^0}^2-q^2)} \int_{s^0_{D}}^\infty dt\frac{\rho_{XD^-}(t,q^2,p^{\prime 2})}{t-p^2}  \nonumber\\
&&+ \frac{1}{(M_{X}^2-p^{\prime2})(M_{D^-}^2-p^2)} \int_{s^0_D}^\infty dt\frac{\rho_{XD^0}(p^2,t,p^{\prime 2})}{t-q^2}\nonumber\\
&&+ \frac{1}{(M_{D^-}^2-p^2)(M_{D^0}^2-q^2)} \int_{s^0_X}^\infty dt\frac{\rho_{XD^-}(p^2,q^2,t)+\rho_{XD^0}(p^2,q^2,t)}{t-p^{\prime2}}+\cdots \, ,
\end{eqnarray}
where $p^\prime=p+q$, the decay constants $f_{\eta_c}$, $f_{\pi}$, $f_{D}$ and  the hadronic coupling constants $G_{X\eta_c\pi}$, $G_{XDD}$ are defined by,
\begin{eqnarray}
\langle0|J_{\eta_c}(0)|\eta_c(p)\rangle&=&\frac{f_{\eta_c}M_{\eta_c}^2}{2m_c} \, , \nonumber \\
\langle0|J_{\pi}(0)|\pi(q)\rangle&=&\frac{f_{\pi}M_{\pi}^2}{m_u+m_d} \, , \nonumber \\
\langle0|J_{D}(0)|D(p/q)\rangle&=&\frac{f_{D}M_{D}^2}{m_c} \, , \nonumber \\
\langle \eta_c(p)\pi(q)|X(p^{\prime})\rangle&=&i G_{X\eta_c\pi} \, , \nonumber\\
\langle D(p)D(q)|X(p^{\prime})\rangle&=&i G_{XDD}   \, .
\end{eqnarray}
The eight functions $\rho_{X\pi}(p^2,t,p^{\prime 2})$, $\rho_{X\eta_c}(t,q^2,p^{\prime 2})$, $\rho_{X\pi}(p^2,q^2,t)$, $\rho_{X\eta_c}(p^2,q^2,t)$, $\rho_{XD^-}(t,q^2,p^{\prime 2})$, $\rho_{XD^0}(p^2,t,p^{\prime 2})$,  $\rho_{XD^-}(p^2,q^2,t)$ and $\rho_{XD^0}(p^2,q^2,t)$ have complex dependence on the transitions
between the ground states and the high resonances  or  continuum states. The definitions of the hadronic coupling constants $G_{X\eta_c\pi}$, $G_{XDD}$ differ from that in Ref.\cite{Wang-Scalar-MPLA}, moreover, in Ref.\cite{Wang-Scalar-MPLA}, an over simplified hadron representation is chosen.

We introduce the notations $C_{X\pi}$, $C_{X\eta_c}$, $C_{X\eta_c}^\prime$, $C_{X\pi}^\prime$, $C_{XD^-}$, $C_{XD^0}$, $C_{XD^-}^\prime$ and $C^\prime_{XD^0}$ to parameterize the net effects,
\begin{eqnarray}
C_{X\pi}&=&\int_{s^0_\pi}^\infty dt\frac{\rho_{X\pi}(p^2,t,p^{\prime 2})}{t-q^2}\, ,\nonumber\\
C_{X\eta_c}&=& \int_{s^0_{\eta_c}}^\infty dt\frac{\rho_{X\eta_c}(t,q^2,p^{\prime 2})}{t-p^2}\, ,  \nonumber\\
C_{X\eta_c}^\prime&=& \int_{s^0_X}^\infty dt\frac{\rho_{X\eta_c}(p^2,q^2,t)}{t-p^{\prime2}} \, , \nonumber\\
C_{X\pi}^\prime&=& \int_{s^0_X}^\infty dt\frac{\rho_{X\pi}(p^2,q^2,t)}{t-p^{\prime2}} \, ,
\end{eqnarray}
\begin{eqnarray}
C_{XD^-}&=& \int_{s^0_{D}}^\infty dt\frac{\rho_{XD^-}(t,q^2,p^{\prime 2})}{t-p^2} \, , \nonumber\\
C_{XD^0}&=& \int_{s^0_D}^\infty dt\frac{\rho_{XD^0}(p^2,t,p^{\prime 2})}{t-q^2}\, ,\nonumber\\
C_{XD^-}^\prime&=& \int_{s^0_X}^\infty dt\frac{\rho_{XD^-}(p^2,q^2,t)}{t-p^{\prime2}} \, ,\nonumber\\
C^\prime_{XD^0}&=& \int_{s^0_X}^\infty dt\frac{\rho^\prime_{XD^0}(p^2,q^2,t)}{t-p^{\prime2}} \, ,
\end{eqnarray}
and rewrite  the correlation functions $\Pi_{1}(p,q)$ and $\Pi_{2}(p,q)$ into the following form,
\begin{eqnarray}
\Pi_{1}(p,q)&=& \frac{f_{\eta_c}M_{\eta_c}^2f_{\pi}M_{\pi}^2\lambda_{X}G_{X\eta_c \pi}}{2m_c(m_u+m_d)} \frac{1}{(M_{X}^2-p^{\prime2})(M_{\eta_c}^2-p^2)(M_{\pi}^2-q^2)}+ \frac{C_{X\pi}}{(M_{X}^2-p^{\prime2})(M_{\eta_c}^2-p^2)} \nonumber\\
&&+ \frac{C_{X\eta_c}}{(M_{X}^2-p^{\prime2})(M_{\pi}^2-q^2)}   + \frac{C_{X\pi}^\prime+C_{X\eta_c}^\prime}{(M_{\eta_c}^2-p^2)(M_{\pi}^2-q^2)} +\cdots \, ,
\end{eqnarray}
\begin{eqnarray}
\Pi_{2}(p,q)&=& \frac{f_{D}^2M_{D}^4 \lambda_{X}G_{XD D}}{4m_c^2} \frac{1}{(M_{X}^2-p^{\prime2})(M_{D^-}^2-p^2)(M_{D^0}^2-q^2)}+ \frac{C_{XD^-}}{(M_{X}^2-p^{\prime2})(M_{D^0}^2-q^2)} \nonumber\\
&&  + \frac{C_{XD^0}}{(M_{X}^2-p^{\prime2})(M_{D^-}^2-p^2)} + \frac{C_{XD^-}^\prime+C_{XD^0}^\prime}{(M_{D^-}^2-p^2)(M_{D^0}^2-q^2)} +\cdots \, .
\end{eqnarray}
In numerical calculations,   we smear  the complex  dependencies of the  $C_{X\pi}$, $C_{X\eta_c}$, $C_{X\eta_c}^\prime$, $C_{X\pi}^\prime$, $C_{XD^-}$, $C_{XD^0}$, $C_{XD^-}^\prime$ and $C^\prime_{XD^0}$  on the variables $p^2,\,p^{\prime 2},\,q^2$, take them as free parameters, and choose the suitable values  to
eliminate the contaminations from the high resonances and continuum states to obtain the stable sum rules with the variations of
the Borel parameters.    In the limit $M_{\pi}^2 \to 0$ and $M_{D^0}^2 \to 0$, we can choose $Q^2=-q^2$ off-shell, and match the terms proportional to $\frac{1}{Q^2}$ at the
hadron side with the ones at the QCD side to obtain QCD sum rules for the momentum dependent hadronic coupling constants $G_{X\eta_c \pi}(Q^2)$ and $G_{XD D}(Q^2)$, then extract the values to the mass-shell $Q^2=-M_{\pi}^2$ or $-M_{D^0}^2$ to obtain the physical values. In fact, the approximations $\frac{1}{M_{D^0}^2-q^2}\approx \frac{1}{Q^2}$ at the hadronic side and $\frac{1}{m_{c}^2-q^2}\approx \frac{1}{Q^2}$ at the QCD side are not good.  We prefer taking the imaginary parts of the correlation functions $\Pi_{1}(p,q)$ and $\Pi_{2}(p,q)$ with respect to $q^2+i\epsilon$ through dispersion relation and obtain the physical spectral densities,  then take Borel transform with respect to the $Q^2$ to obtain the QCD sum rules for the physical hadronic coupling constants.

We have to be cautious in matching the QCD side with the hadronic side of the correlation functions $\Pi_{1}(p,q)$ and $\Pi_{2}(p,q)$, as there appears  the variable $p^{\prime2}=(p+q)^2$.
We rewrite the correlation functions $\Pi_{1}(p,q)$ and $\Pi_{2}(p,q)$ at the hadronic side into the following form through dispersion relation,
\begin{eqnarray}
\Pi_1(p,q)&=&\Pi^H_1(p^{\prime 2},p^2,q^2)\nonumber\\
&=&\int_{(M_{\eta_c}+M_{\pi})^2}^{s_X^0}ds^\prime \int_{4m_c^2}^{s^0_{\eta_c}}ds \int_0^{u^0_{\pi}}du  \frac{\rho_H^1(s^\prime,s,u)}{(s^\prime-p^{\prime2})(s-p^2)(u-q^2)}+\cdots\, , \\
\Pi_2(p,q)&=&\Pi^H_2(p^{\prime 2},p^2,q^2)\nonumber\\
&=&\int_{4M_{D}^2}^{s^0_X}ds^\prime \int_{m_c^2}^{s^0_D}ds \int_{m_c^2}^{u^0_D}du  \frac{\rho_H^2(s^\prime,s,u)}{(s^\prime-p^{\prime2})(s-p^2)(u-q^2)}+\cdots\, ,
\end{eqnarray}
where the $\rho_H^1(s^\prime,s,u)$ and $\rho_H^2(s^\prime,s,u)$ are the hadronic spectral densities,
\begin{eqnarray}
\rho_H^1(s^\prime,s,u)&=&{\lim_{\epsilon_3\to 0}}\,\,{\lim_{\epsilon_2\to 0}} \,\,{\lim_{\epsilon_1\to 0}}\,\,\frac{ {\rm Im}_{s^\prime}\, {\rm Im}_{s}\,{\rm Im}_{u}\,\Pi^H_1(s^\prime+i\epsilon_3,s+i\epsilon_2,u+i\epsilon_1) }{\pi^3} \, ,\\
\rho_H^2(s^\prime,s,u)&=&{\lim_{\epsilon_3\to 0}}\,\,{\lim_{\epsilon_2\to 0}} \,\,{\lim_{\epsilon_1\to 0}}\,\,\frac{ {\rm Im}_{s^\prime}\, {\rm Im}_{s}\,{\rm Im}_{u}\,\Pi^H_2(s^\prime+i\epsilon_3,s+i\epsilon_2,u+i\epsilon_1) }{\pi^3} \, .
\end{eqnarray}
The ground state masses have the relations $M_{X}>M_{\eta_c({\rm 2S})}>M_{\eta_c}\gg M_{\pi}$ and $M_{X}\approx 2 M_{D}$, while the continuum threshold parameters have the relations $\sqrt{s^0_X}\approx \sqrt{s^0_{\eta_c}}+\sqrt{u^0_{\pi}}$, $\sqrt{s^0_X}> \sqrt{s^0_{\eta_c}}\gg \sqrt{u^0_{\pi}}$,  $\sqrt{s^0_X}\approx \sqrt{s^0_{D}}+\sqrt{u^0_{D}}-0.6\,\rm{GeV}$ and $s^0_D=u^0_D$ \cite{Wang-4200,WangJHEP}.

  Now we  set $s^0_{\eta_c}=s^0_X$, $p^{\prime2}=p^2$ and carry out the integral over $ds^\prime$, the contribution of the $\eta_c(2\rm S)$ is included in, we have to take into account the contribution of the  $\rho_{X\eta_c}(t,q^2,p^{\prime 2})$ explicitly. On the other hand, we set $\sqrt{s^0_X}= \sqrt{s^0_{D}}+\sqrt{u^0_{D}}$, $p^{\prime2}=4p^2$ and carry out the integral over $ds^\prime$, the contribution of the $X(2\rm S)$ is included in, we have to take into account the contribution of the $\rho_{XD^-}(p^2,q^2,t)$ explicitly. The pole terms below the continuum thresholds $s^0_X$, $s^0_{\eta_c}$, $u^0_{\pi}$, $s^0_{D}$ and $u^0_{D}$ can be  written as
\begin{eqnarray}
\Pi_{1}(p,q)&=& \frac{f_{\eta_c}M_{\eta_c}^2f_{\pi}M_{\pi}^2\lambda_{X}G_{X\eta_c \pi}}{2m_c(m_u+m_d)} \frac{1}{(M_{X}^2-p^{2})(M_{\eta_c}^2-p^2)(M_{\pi}^2-q^2)}+ \frac{C_{X\eta_c}}{(M_{X}^2-p^{2})(M_{\pi}^2-q^2)}  \, ,  \nonumber\\
\end{eqnarray}
\begin{eqnarray}
\Pi_{2}(p,q)&=& \frac{f_{D}^2M_{D}^4 \lambda_{X}G_{XD D}}{16m_c^2} \frac{1}{(\widetilde{M}_{X}^2-p^{2})(M_{D^-}^2-p^2)(M_{D^0}^2-q^2)}  + \frac{C_{XD^-}^\prime }{(M_{D^-}^2-p^2)(M_{D^0}^2-q^2)}   \, , \nonumber\\
\end{eqnarray}
where $\widetilde{M}_{X}^2=\frac{M_X^2}{4}$.

We carry out the operator product expansion up to the vacuum condensates of dimension 5 and neglect the tiny contribution of the gluon condensate.
In this article, we take into account both the connected and disconnected Feynman diagrams, just like in the QCD sum rules for the two-body strong decays of the $Z_c(4200)$ and $X(5568)$  \cite{Wang-4200,Zhu-4200-Wang-5568},   which is contrary to Ref.\cite{Nielsen3900}, where only the connected Feynman diagrams are taken into account to study the width of the $Z_c(3900)$. In Ref.\cite{Wang-Scalar-MPLA}, we only take into account the connected Feynman diagrams in calculating the width of the lowest scalar hidden-charm tetraquark state and obtain the value $\Gamma\approx 21\,\rm{MeV}$.

In calculations, we observe that there appears  $\frac{q\cdot p}{q^2}$ in the terms associated with the $\langle\bar{q}q\rangle$ and $\langle\bar{q}g_s\sigma Gq\rangle$ in the correlation function $\Pi_{1}(p,q)$, which disappears  after performing the Borel transform  with respect to the variable $Q^2=-q^2$, as $\frac{q\cdot p}{q^2}=\frac{p^{\prime2}-p^2-q^2}{2q^2}=-\frac{1}{2}$ by setting $p^{\prime2}=p^2=-P^2$.

Once the analytical expressions of the  correlation functions $\Pi_{1}(p,q)$ and $\Pi_{2}(p,q)$ at the QCD level are gotten, we can
 obtain the QCD spectral densities through dispersion relation, take the quark-hadron  duality below the continuum thresholds, then we set $p^{\prime2}=p^2$ and $p^{\prime2}=4p^2$ for the correlation functions  $\Pi_{1}(p,q)$ and $\Pi_{2}(p,q)$ respectively,  and take  the double Borel transforms with respect to the variables    $P^2=-p^2 $ and $Q^2=-q^2$ respectively to obtain the following QCD sum rules,
\begin{eqnarray}
&&\frac{f_{\eta_c}M_{\eta_c}^2f_{\pi}M_{\pi}^2\lambda_{X}G_{X\eta_c \pi}}{2m_c(m_u+m_d)}\frac{1}{M_{X}^2-M_{\eta_c}^2} \left[ \exp\left(-\frac{M_{\eta_c}^2}{T^2} \right)-\exp\left(-\frac{M_{X}^2}{T^2} \right)\right]\exp\left(-\frac{M_{\pi}^2}{T_2^2} \right) \nonumber\\
&&+C_{X\eta_c} \exp\left(-\frac{M_{X}^2}{T^2} -\frac{M_{\pi}^2}{T_2^2} \right)=\frac{3}{128\pi^4}\int_{4m_c^2}^{s^0_{X}} ds \int_{0}^{u^0_{\pi}} du  \,su\, \sqrt{1-\frac{4m_c^2}{s}}\exp\left(-\frac{s}{T^2} -\frac{u}{T_2^2} \right)\, , \nonumber\\
\end{eqnarray}
\begin{eqnarray}
&&\frac{f_{D}^2M_{D}^4 \lambda_{X}G_{XD D}}{16m_c^2} \frac{1}{\widetilde{M}_{X}^2-M_{D}^2}\left[ \exp\left(-\frac{M_{D}^2}{T^2} \right)-\exp\left(-\frac{\widetilde{M}_{X}^2}{T^2} \right)\right]\exp\left(-\frac{M_{D}^2}{T_2^2} \right) \nonumber\\
&&+C_{XD^-}^\prime  \exp\left(-\frac{M_{D}^2}{T^2} -\frac{M_{D}^2}{T_2^2} \right)=-\frac{3}{256\pi^4}\int_{m_c^2}^{s^0_{D}} ds \int_{m_c^2}^{u^0_{D}} du  \frac{(s-m_c^2)^2(u-m_c^2)^2\left[(3s-u)m_c^2+2su \right]}{s^2u^2}\nonumber\\
&&\exp\left(-\frac{s}{T^2} -\frac{u}{T_2^2} \right)+\frac{m_c\langle\bar{q}q\rangle}{32\pi^2} \int_{m_c^2}^{u^0_{D}} du \frac{(u-m_c^2)^2(u+3m_c^2)}{u^2} \exp\left(-\frac{m_c^2}{T^2} -\frac{u}{T_2^2} \right) \nonumber\\
&&+\frac{m_c\langle\bar{q}q\rangle}{32\pi^2} \int_{m_c^2}^{s^0_{D}} ds \frac{(s-m_c^2)^2(m_c^2-5s)}{s^2}\exp\left(-\frac{s}{T^2} -\frac{m_c^2}{T_2^2} \right) \nonumber\\
&&+\frac{m_c\langle\bar{q}g_s\sigma Gq\rangle}{128\pi^2} \int_{m_c^2}^{s^0_{D}} ds \frac{10s^2-7sm_c^2+m_c^4}{s^2}\exp\left(-\frac{s}{T^2} -\frac{m_c^2}{T_2^2} \right) \nonumber\\
&&+\frac{m_c\langle\bar{q}g_s\sigma Gq\rangle}{128\pi^2} \int_{m_c^2}^{u^0_{D}} du \frac{2u^2+5um_c^2-3m_c^4}{u^2}\exp\left(-\frac{m_c^2}{T^2} -\frac{u}{T_2^2} \right) \, ,
\end{eqnarray}
where the $T^2$ and $T_2^2$ are the Borel parameters.
In the two QCD sum rules, the terms depend on $T^2_2$ can be factorized out explicitly,
\begin{eqnarray}
&&\frac{f_{\eta_c}M_{\eta_c}^2f_{\pi}M_{\pi}^2\lambda_{X}G_{X\eta_c \pi}}{2m_c(m_u+m_d)}\frac{1}{M_{X}^2-M_{\eta_c}^2} \left[ \exp\left(-\frac{M_{\eta_c}^2}{T^2} \right)-\exp\left(-\frac{M_{X}^2}{T^2} \right)\right] \nonumber\\
&&+C_{X\eta_c} \exp\left(-\frac{M_{X}^2}{T^2} \right)=\frac{3}{128\pi^4}\int_{4m_c^2}^{s^0_{X}} ds \int_{0}^{u^0_{\pi}} du  \,su\, \sqrt{1-\frac{4m_c^2}{s}}\exp\left(-\frac{s}{T^2} -\frac{u-M_{\pi}^2}{T_2^2} \right)\, , \nonumber\\
\end{eqnarray}
\begin{eqnarray}
&&\frac{f_{D}^2M_{D}^4 \lambda_{X}G_{XD D}}{16m_c^2} \frac{1}{\widetilde{M}_{X}^2-M_{D}^2}\left[ \exp\left(-\frac{M_{D}^2}{T^2} \right)-\exp\left(-\frac{\widetilde{M}_{X}^2}{T^2} \right)\right] \nonumber\\
&&+C_{XD^-}^\prime  \exp\left(-\frac{M_{D}^2}{T^2}  \right)=-\frac{3}{256\pi^4}\int_{m_c^2}^{s^0_{D}} ds \int_{m_c^2}^{u^0_{D}} du  \frac{(s-m_c^2)^2(u-m_c^2)^2\left[(3s-u)m_c^2+2su \right]}{s^2u^2}\nonumber\\
&&\exp\left(-\frac{s}{T^2} -\frac{u-M_{D}^2}{T_2^2} \right)+\frac{m_c\langle\bar{q}q\rangle}{32\pi^2} \int_{m_c^2}^{u^0_{D}} du \frac{(u-m_c^2)^2(u+3m_c^2)}{u^2} \exp\left(-\frac{m_c^2}{T^2} -\frac{u-M_{D}^2}{T_2^2} \right) \nonumber\\
&&+\frac{m_c\langle\bar{q}q\rangle}{32\pi^2} \int_{m_c^2}^{s^0_{D}} ds \frac{(s-m_c^2)^2(m_c^2-5s)}{s^2}\exp\left(-\frac{s}{T^2} -\frac{m_c^2-M_{D}^2}{T_2^2} \right) \nonumber\\
&&+\frac{m_c\langle\bar{q}g_s\sigma Gq\rangle}{128\pi^2} \int_{m_c^2}^{s^0_{D}} ds \frac{10s^2-7sm_c^2+m_c^4}{s^2}\exp\left(-\frac{s}{T^2} -\frac{m_c^2-M_{D}^2}{T_2^2} \right) \nonumber\\
&&+\frac{m_c\langle\bar{q}g_s\sigma Gq\rangle}{128\pi^2} \int_{m_c^2}^{u^0_{D}} du \frac{2u^2+5um_c^2-3m_c^4}{u^2}\exp\left(-\frac{m_c^2}{T^2} -\frac{u-M_{D}^2}{T_2^2} \right) \, ,
\end{eqnarray}
the dependence on the $T^2_2$ is rather trivial, $\exp\left(-\frac{u-M_{\pi}^2}{T_2^2} \right)$, $\exp\left(-\frac{u-M_{D}^2}{T_2^2} \right)$,  $\exp\left(-\frac{m_c^2-M_{D}^2}{T_2^2} \right)$, which differ from the QCD sum rules for the three-meson hadronic coupling constants greatly \cite{Nielsen-PPNP}. It is difficult to obtain $T_2^2$ independent regions in the present QCD sum rules, as no other terms to stabilize the QCD sum rules.
We can take the local limit $T^2_2\to\infty$, which is so called  local-duality limit (the local QCD sum rules are reproduced from the   original
QCD sum rules in infinite Borel parameter limit) \cite{Local-QCDSR}, then $\exp\left(-\frac{u}{T_2^2} \right)=\exp\left(-\frac{m_c^2}{T_2^2} \right)=\exp\left(-\frac{M_\pi^2}{T_2^2} \right)=\exp\left(-\frac{M_D^2}{T_2^2} \right)=1$, the two QCD sum rules are greatly simplified.

The hadronic input parameters are chosen  as   $M_{\eta_c}=2.9836\,\rm{GeV}$,  $f_{\pi}=0.130\,\rm{GeV}$ \cite{PDG}, $\sqrt{s^0_{\pi}}=0.85\,\rm{GeV}$ \cite{Wang-4200}, $M_{D}=1.87\,\rm{GeV}$, $f_{D}=208\,\rm{MeV}$, $s^0_{D}=u^0_{D}=6.2\,\rm{GeV}^2$  \cite{WangJHEP},
 $f_{\eta_c}=0.387 \,\rm{GeV}$  \cite{Becirevic},  $\sqrt{s^0_{X}}=4.4\,\rm{GeV}$,  $M_X=3.86\,\rm{GeV}$, $\lambda_{X}=2.02\times 10^{-2}\,\rm{GeV}^5$ (this work), and $f_{\pi}M^2_{\pi}/(m_u+m_d)=-2\langle \bar{q}q\rangle/f_{\pi}$ from the Gell-Mann-Oakes-Renner relation.
The unknown parameters are chosen as $C_{X\eta_c}=0.0063\,\rm{GeV}^8 $ and $C_{XD^-}^\prime=-0.0071\,\rm{GeV}^8 $  to obtain  platforms in the Borel windows $T^2=(2.5-2.9)\,\rm{GeV}^2$ (this work) and $T^2=(1.3-1.7)\,\rm{GeV}^2$ \cite{WangJHEP}, respectively. The input parameters at the QCD side are chosen as the same ones in the two-point QCD sum rules for the $X^*(3860)$.
Then it is easy to obtain the values of the hadronic coupling constants,
\begin{eqnarray}
G_{X\eta_c \pi} &=&1.28\pm0.18\,\rm{GeV}\, , \nonumber\\
|G_{XDD}|&=&12.3\pm4.5\,\rm{GeV}\, .
\end{eqnarray}
 In Fig.4, we plot the hadronic coupling constants $G_{X\eta_c \pi}$ and $G_{XDD}$ at much larger intervals than the Borel windows. From the figure, we can see
 that the values of the hadronic coupling constants $G_{X\eta_c \pi}$ and $G_{XDD}$ are rather stable with variations of the Borel parameters, so we expect to make reliable predictions.
     The uncertainties of the $G_{X\eta_c \pi}$ and $G_{XDD}$ lead to the uncertainties $\delta\Gamma(X^*(3860)\to \eta_c\pi^-)/\Gamma(X^*(3860)\to \eta_c\pi^-)=2\delta G_{X\eta_c \pi}/G_{X\eta_c \pi}=28\%$ and $\delta\Gamma(X^*(3860)\to D^-D^0)/\Gamma(X^*(3860)\to D^-D^0)=2\delta G_{XDD}/G_{XDD}= 73\%$.

 \begin{figure}
 \centering
 \includegraphics[totalheight=5cm,width=7cm]{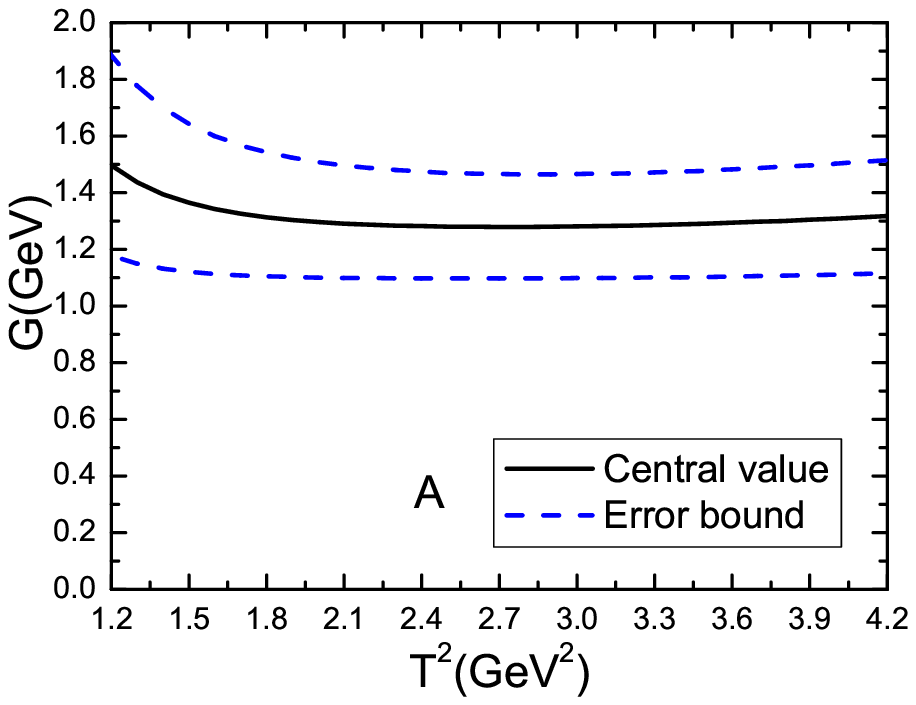}
 \includegraphics[totalheight=5cm,width=7cm]{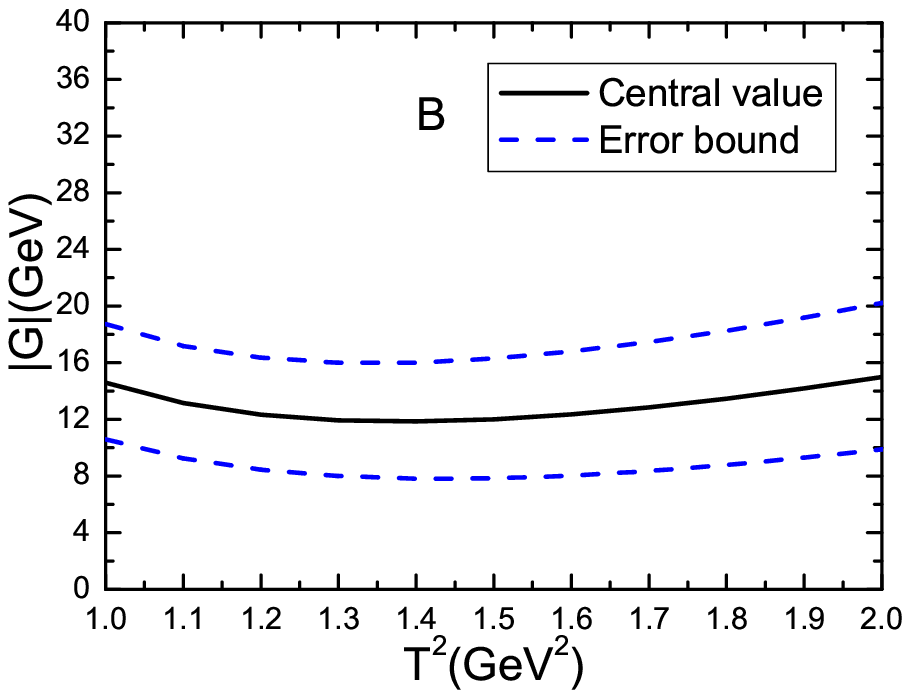}
        \caption{ The hadronic coupling constants  with variations of the Borel parameters $T^2$, where the $A$ and $B$ correspond to the $G_{X\eta_c\pi}$ and $G_{XDD}$, respectively.  }
\end{figure}

We choose the masses $M_{X}=3.862\,\rm{GeV}$ \cite{Belle-3860}, $M_{\eta_c}=2.9836\,\rm{GeV}$, $M_{\pi}=0.13957\,\rm{GeV}$, $M_{D^-}=1.8695\,\rm{GeV}$, $M_{D^0}=1.8649\,\rm{GeV}$ \cite{PDG},  and obtain the numerical  values of the   decay widths,
\begin{eqnarray}
\Gamma(X^*(3860)\to \eta_c\pi^-)&=& \frac{G_{X\eta_c \pi}^2 \,p_{\eta_c\pi}}{8\pi M_X^2}\nonumber\\
&=& 3.4\pm1.0\,\rm{MeV}   \, ,\nonumber\\
\Gamma(X^*(3860)\to D^-D^0)&=&\frac{G_{XDD}^2\, p_{DD}}{8\pi M_X^2}\nonumber\\
&=&198.7\pm 145.1 \,\rm{MeV}  \, ,
\end{eqnarray}
where
\begin{eqnarray}
p_{\eta_c\pi}&=&\frac{\sqrt{\left[ M_X^2-(M_{\eta_c}+M_{\pi})^2\right]\left[ M_X^2-(M_{\eta_c}-M_{\pi})^2\right]}}{2M_X}\, , \nonumber\\
p_{DD}&=&\frac{\sqrt{\left[ M_X^2-(M_{D^0}+M_{D^-})^2\right]\left[ M_X^2-(M_{D^0}-M_{D^-})^2\right]}}{2M_X}\, .
\end{eqnarray}
If we saturate the width of the $X^*(3860)$ with the strong decays to the meson pairs $\eta_c\pi^-$ and $D^-D^0$, then
$\Gamma_{X}= 202\pm 146 \,\rm{MeV}$, which is in excellent agreement with the experimental value $\Gamma_{X} = 201^{+154}_{-67}{}^{+88}_{-82}\,\rm{MeV}$ from the Belle collaboration \cite{Belle-3860}, the present calculations support  assigning the $X^*(3860)$ to be  the $C\gamma_5\otimes \gamma_5C$ type hidden-charm   tetraquark  state.

\section{Conclusion}
In this article, we tentatively assign the $X^*(3860)$ to be  the $C\gamma_5\otimes \gamma_5C$ type scalar  tetraquark  state,   study its mass and width with the QCD sum rules, special attention is paid to calculating the hadronic coupling constants $G_{X\eta_c\pi}$ and $G_{XDD}$. We obtain the values $M_{X}=3.86 \pm 0.09\,\rm{GeV}$ and  $\Gamma_{X}= 202\pm146 \,\rm{MeV}$, which are consistent with the experimental data $M_{X}=3862^{+26}_{-32}{}^{+40}_{-13}\,\rm{MeV}$  and $\Gamma_{X}=201^{+154}_{-67}{}^{+88}_{-82}\,\rm{MeV}$, respectively. The dominant decay mode of the neutral  partner $X^{*0}(3860)$ is $X^{*0}(3860)\to D \bar{D}$, which is also consistent with the fact that the $X^{*0}(3860)$ is observed in the process $e^+ e^- \rightarrow
J/\psi D \bar{D}$.  The present work  supports assigning  the   $X^*(3860)$ to be  the $C\gamma_5\otimes \gamma_5C$ type hidden-charm  tetraquark  state.

\section*{Acknowledgements}
This  work is supported by National Natural Science Foundation, Grant Number 11375063.


\begin{thebibliography}{99}

\bibitem{Belle-3860} K. Chilikin et al,   Phys. Rev. {\bf D95} (2017) 112003.

\bibitem{GI-model}  T. Barnes, S. Godfrey and E. S. Swanson, Phys. Rev. {\bf D72} (2005) 054026.

\bibitem{ChaoKT} B. Q. Li and K. T. Chao, Phys. Rev. {\bf D79} (2009) 094004.



\bibitem{Belle2004} S. K. Choi  et al,  Phys. Rev. Lett. {\bf 94} (2005) 182002.

\bibitem{BaBar2007} B. Aubert et al, Phys. Rev. Lett. {\bf 101} (2008) 082001.

\bibitem{Belle2010} S. Uehara  et al,  Phys. Rev. Lett. {\bf 104} (2010) 092001.


\bibitem{Lebed-3915} R. F. Lebed and A. D. Polosa, Phys. Rev. {\bf D93} (2016) 094024.

\bibitem{Wang-3915-CgmCgm} Z. G. Wang, Eur. Phys. J. {\bf C77} (2017) 78.

\bibitem{Wang-3915-C5C5}  Z. G. Wang, Eur. Phys. J. {\bf A53} (2017) 19.

\bibitem{WangScalar-2009} Z. G. Wang, Phys. Rev. {\bf D79} (2009) 094027;
Z. G. Wang,  Eur. Phys. J. {\bf C67} (2010) 411.


\bibitem{Wang-Axial-V-tetraquark} Z. G. Wang, Eur. Phys. J. {\bf C70} (2010) 139.

\bibitem{Wang-4660-2014} Z. G. Wang, Eur. Phys. J. {\bf C74} (2014)  2874.

\bibitem{Wang-Scalar-MPLA} Z. G. Wang, Mod. Phys. Lett. {\bf A29} (2014) 1450207.


\bibitem{Wang-1601} Z. G. Wang,  Eur. Phys. J. {\bf C76} (2016)  387.


\bibitem{Wang-4200} Z. G. Wang, Int. J. Mod. Phys. {\bf A30} (2015) 1550168.


\bibitem{Zhu-4200-Wang-5568} W. Chen, T. G. Steele, H. X. Chen and S. L. Zhu, Eur. Phys. J. {\bf C75} (2015)  358;
Z. G. Wang, Eur. Phys. J. {\bf C76} (2016)  279.

\bibitem{Nielsen3900} J. M. Dias, F. S. Navarra, M. Nielsen and C. M. Zanetti, Phys. Rev. {\bf D88} (2013) 016004.

\bibitem{SVZ79} M. A. Shifman, A. I. Vainshtein and V. I. Zakharov, Nucl. Phys. {\bf B147} (1979) 385; Nucl. Phys. {\bf B147} (1979) 448.

\bibitem{Reinders85} L. J. Reinders, H. Rubinstein and S. Yazaki, Phys. Rept. {\bf 127} (1985) 1.

\bibitem{Colangelo-Review}  P. Colangelo and A. Khodjamirian, hep-ph/0010175.

\bibitem{PDG}   K. A. Olive et al, Chin. Phys. {\bf C38} (2014) 090001.

\bibitem{Wang4430} Z. G. Wang,  Commun. Theor. Phys. {\bf 63} (2015) 325.


\bibitem{WangHuangTao-3900} Z. G. Wang and T. Huang,  Phys. Rev. {\bf D89} (2014) 054019.

\bibitem{Wang-Huang-NPA-2014} Z. G. Wang and T. Huang, Nucl. Phys. {\bf A930} (2014) 63.



\bibitem{Belle-X3940}  K. Abe et al, Phys. Rev. Lett. {\bf 98} (2007) 082001;
P. Pakhlov et al, Phys. Rev. Lett. {\bf 100} (2008) 202001.

\bibitem{Azizi-2015} K. Azizi, Y. Sarac and H. Sundu, Phys. Rev. {\bf D90} (2014)  114011;
K. Azizi, Y. Sarac and H. Sundu, Nucl. Phys. {\bf A943} (2015) 159.

\bibitem{Azizi-2015-YGL} K. Azizi, Y. Sarac and H. Sundu, Phys. Rev. {\bf D92} (2015)  014022.


\bibitem{WangPRD-2014} Z. G. Wang,  Phys. Rev. {\bf D89} (2014)  034017;
Z. G. Wang, Eur. Phys. J. {\bf C74} (2014)  3123.


\bibitem{Nielsen-PPNP}  M. E. Bracco, M. Chiapparini, F. S. Navarra and M. Nielsen, Prog. Part. Nucl. Phys. {\bf 67} (2012) 1019.

\bibitem{Local-QCDSR} V. A. Nesterenko and A. V. Radyushkin, Phys. Lett. {\bf 115B} (1982) 410;
A. V. Radyushkin, Acta Phys. Polon. {\bf B26} (1995) 2067;
A. P. Bakulev,  Nucl. Phys. Proc. Suppl. {\bf 198} (2010) 204.


\bibitem{WangJHEP}  Z. G. Wang, JHEP {\bf 1310} (2013) 208;
Z. G. Wang, Eur. Phys. J. {\bf C75} (2015) 427.


\bibitem{Becirevic} D. Becirevic, G. Duplancic, B. Klajn, B. Melic and F. Sanfilippo,  Nucl. Phys. {\bf B883} (2014) 306.


\end{thebibliography}
\end{document}